\documentclass[preprint,nofootinbib]{revtex4-1}%
\usepackage{amssymb}
\usepackage{amsfonts}
\usepackage{amsmath}
\usepackage{graphicx}
\usepackage[usenames]{color}%
\providecommand{\U}[1]{\protect\rule{.1in}{.1in}}
\definecolor{blue}{rgb}{0,0,1}

\definecolor{red}{rgb}{1,0,0}

\begin{document}

\title{Asymptotically locally AdS and flat black holes in the presence of an electric
field in the Horndeski scenario}
\author{Adolfo Cisterna}
\email{adolfo.cisterna.r@mail.pucv.cl}
\affiliation{Instituto de F\'{\i}sica, Pontificia Universidad Cat\'{o}lica de
Valpara\'{\i}so, Av. Universidad 330, Curauma, Valpara\'{\i}so, Chile.}
\author{Cristi\'an Erices}
\email{erices@cecs.cl}
\affiliation{Departamento de F\'isica, Universidad de Concepci\'on, Casilla, 160-C, Concepci\'on, Chile.}
\affiliation{Centro de Estudios Cient\'{\i}ficos (CECs), Casilla 1469, Valdivia, Chile.}

\begin{abstract}
Asymptotically locally AdS and asymptotically flat black hole
solutions are found for a particular case of the Horndeski action. The action
contains the Einstein-Hilbert term with a cosmological constant, a real scalar field
with a non minimal kinetic coupling given by the Einstein tensor, the minimal kinetic coupling and the Maxwell term. There is no scalar potential.
The solution has two integration constants related with the mass and the
electric charge. The solution is given for all dimensions. A new
class of asymptotically locally flat spherically symmetric black holes is
found when the minimal kinetic coupling vanishes and the cosmological constant is present. In this case we get a solution
which represents an electric Universe. The electric field at
infinity is only supported by $\Lambda$. When the cosmological constant vanishes the black hole is asymptotically flat.
\end{abstract}
\date{January 17, 2014}
\maketitle

\section{Introduction}

Scalar fields have a prominent role in high energy physics. At subatomic
scales they are an essential part of the quantum description of the
electroweak interaction. Indeed, a foundamental scalar field excitation is
given by the well known Brout-Englert-Higgs particle, which allows a
consistent mathematical description of the the short range of the weak force
and lepton masses.

At the galactic and cosmological scales, scalar fields arise as the simplest
candidate to the explanation of many phenomena. At these scales, the general
theory of relativity successfully describes the gravitational interaction.
However, despite the great success of the theory, it cannot give a
satisfactory description of certain cosmological phenomena, such as the origin
of the early Universe and its late time accelerated expansion, as well as the
presence of dark matter and dark energy. The properties of these phenomena
make the scalar field a suitable candidate able to solve such unknowns, giving
rise to a wide variety of theories such as Brans-Dicke theory \cite{BD},
inflation theories and several cosmological models \cite{dancosm}\cite{chungcosm}\cite{macorracosm}.

Moreover scalar fields appear naturally in theories like Kaluza-Klein
compactifications and in theories that intend to give a natural description
of gravity at the quantum level, such as string theory, which includes the
dilaton scalar field.

While it is true that the study of scalar-tensor theories is not a new topic,
currently a great interest resurfaced due to the study of galileon theories
and their applications. This have revived the study of the most general
scalar-tensor theory which has second order field equations and second order
energy-momentum tensor, problem that was solved by Horndeski four decades ago
\cite{Horn1}. Horndeski theory along with a big amount of interesting
properties also includes galileon gravity \cite{rinaldi3} and massive gravity
\cite{rinaldi4}.

If we focus our attention in a four dimensional curved spacetime, the most
general Lagrangian which can be constructed with the above properties is given
by
\begin{equation}
L=\lambda_{1}\delta_{efhi}^{abdc}R_{ab}^{ef}R_{cd}^{hi}+\lambda_{2}%
\delta_{def}^{abc}\nabla_{a}\phi\nabla^{d}\phi R_{cd}^{ef}+\lambda_{3}%
\delta_{cd}^{ab}R_{ab}^{cd}+\Theta+B\epsilon^{abdc}R^{p}{}_{qab}R^{q}{}_{pcd}
\label{horn4}%
\end{equation}
where $B$ is a constant, $\lambda_{i}$ are arbitrary functions of the scalar
$\phi$ and $\Theta$ is and arbitrary function of the scalar field and its
squared gradient, i.e. $\Theta=\Theta(\nabla_{a}\phi\nabla^{a}\phi,\phi)$.

At this point, we can see that obtaining scalar field Lagrangians, whose kinetic
term has non-minimal couplings with the curvature, is possible. In a
cosmological context, theories where this non-minimal derivative coupling is
given by the Einstein tensor, provides an expansion of the Universe without a
scalar potential \cite{rinaldi6}. Accelerating behaviors were observed as well
in the case of a coupling given by the Ricci tensor \cite{rinaldi7}. Many
models appeared in this context \cite{rinaldia}\cite{rinaldib}\cite{rinaldic}.

Let us focus our attention on kinetic terms $S$ which are quadratic in the
derivatives of the field in arbitrary dimension $n$. Requiring second order
energy-momentum tensor, as well as field equations for the field, single out $S$
as a linear combination of the following terms
\begin{equation}
S^{\left(  p\right)  }=E_{\mu\nu}^{(p)}\nabla^{\mu}\phi\nabla^{\nu}\phi\ ,
\label{familyn}%
\end{equation}
where $E_{\mu\nu}^{(p)}$ is $p$-$th$ order Lovelock tensors\footnote{The most
general symmetric tensors which are divergency-free and contain up to second
order derivatives of the metric.}\cite{Lovelock}
\begin{equation}
E^{(p)}{}_{\mu}^{\nu}=\delta_{\mu\beta_{1}...\beta_{2p}}^{\nu\alpha
_{1}...\alpha_{2p}}R_{\ \ \alpha_{1}\alpha_{2}}^{\beta_{1}\beta_{2}%
}...R_{\ \ \alpha_{2p-1}\alpha_{2p}}^{\beta_{2p-1}\beta_{2p}}\ ,
\end{equation}

By setting $p=0$, the standard kinetic term is therefore obtained. Since
$E_{\mu\nu}^{\left(  1\right)  }$ is proportional to the Einstein tensor, the
first non-standard term in (\ref{familyn}) already includes a non-minimal
kinetic coupling of the scalar field and the curvature.

In this paper we shall focus on the study of black hole solutions and their properties that emerge
from this theory. The action principle is given by
\begin{equation}
I[g_{\mu\nu},\phi]=\int\sqrt{-g}d^{n}x\ \left[  \kappa\left(  R-2\Lambda
\right)  -\frac{1}{2}\left(  \alpha g_{\mu\nu}-\eta G_{\mu\nu}\right)
\nabla^{\mu}\phi\nabla^{\nu}\phi-\frac{1}{4}F_{\mu\nu}F^{\mu\nu}\right]  \ .
\label{action}%
\end{equation}
The strength of the non minimal kinetic coupling is controlled by $\eta$. Here $\kappa:=\frac{1}{16\pi G}$. The
possible values of the dimensionfull parameters $\alpha$ and $\eta$ will be
determined below requiring the positivity of the energy density of the
matter field.

The first exact black hole solution to this system was found by Rinaldi in
\cite{rinaldisolo} for the case of vanishing cosmological constant
$\Lambda$ and without the Maxwell term. In that solution the scalar field
becomes imaginary in the domain of outer communications, and the weak energy
condition is violated outside of the horizon.

A great interest has been generated by spacetimes which are asymptotically of constant curvature, particularly asymptotically AdS spacetimes. This interest is largely motivated by
the AdS/CFT correspondence \cite{malda} which relates the observables in a gauged supergravity theory with those of a conformal field theory in one dimension less. In this way, black
hole solutions with a negative cosmological constant are important because in
principle they could provide the possibility of studying the phase diagram of a CFT theory. As we know, a black hole in an asymptotically flat spacetime is
thermodynamically unstable. In order to solve this problem it is possible to
put the black hole inside a cavity of finite size. However, there is an alternative method to
stabilize such a black hole. It consists in adding a negative cosmological
constant. The
properties of the AdS spacetime stabilize the black hole simulating a
reflecting cavity.

Therefore, it seems natural to study the case where a negative cosmological
constant is present. This was done in \cite{ado}, where a real scalar field
outside the horizon was found and where the positivity of the energy density
is given by this reality condition. Recently in reference \cite{charm} it has
been shown that allowing the scalar to depend on time permits to construct a
black hole solution in which the scalar field is analytic at the future or at
the past horizon. In a similar context exact solutions were found in
\cite{portu}.

The aim of this work is to continue in this line and generalize the results in
reference \cite{ado} by adding a Maxwell term given by a spherically symmetric
gauge field $A=A_{0}(r)dt$.

A numerical solution in this case was found in \cite{NumericalSol}, where
phase transitions to charged black hole with complex anisotropic scalar hair
were explored. We also extend the solution to the topological case in
arbitrary dimension $n\geq4$ and show that it is also possible to obtain a
non-trivial solution when $\alpha=0$. In this later case, when the black hole is
spherically symmetric, we obtain an asymptotically locally flat black hole with
$\Lambda\neq0$ and an asymptotically flat black hole (i.e the metric is
Minkowski at spatial infinity\footnote{This is the difference with the
asymptotically \textit{locally} flat solution. However, both solutions have
curvatures which vanishes at spatial infinity.}) when $\Lambda=0$.

The variation of the action (\ref{action}) with respect to the metric tensor,
the scalar field and the gauge field yields
\begin{equation}
G_{\mu\nu}+\Lambda g_{\mu\nu}=\frac{\alpha}{2\kappa}T_{\mu\nu}^{(1)}%
+\frac{\eta}{2\kappa}T_{\mu\nu}^{\left(  2\right)  }+\frac{1}{2\kappa}%
T_{\mu\nu}^{em}\ , \label{eqmetric}%
\end{equation}%
\begin{equation}
\nabla_{\mu}\left[  \left(  \alpha g^{\mu\nu}-\eta G^{\mu\nu}\right)
\nabla_{\nu}\phi\right]  =0\ , \label{eqphi}%
\end{equation}%
\begin{equation}
\nabla_{\mu}F^{\mu\nu}=0\ ,
\end{equation}
respectively. Here we have defined\footnote{We use a normalized
symmetrization $A_{(\mu\nu)}:=\frac{1}{2}\left(  A_{\mu\nu}+A_{\nu\mu}\right)
$.} 
\begin{eqnarray*}
T_{\mu\nu}^{\left(  1\right)  } &=&\nabla_{\mu}\phi\nabla_{\nu}\phi-\frac
{1}{2}g_{\mu\nu}\nabla_{\lambda}\phi\nabla^{\lambda}\phi\ ,\\
T_{\mu\nu}^{\left(  2\right)  }&=&\frac{1}{2}\nabla_{\mu}\phi\nabla_{\nu
}\phi R-2\nabla_{\lambda}\phi\nabla_{(\mu}\phi R_{\nu)}^{\lambda}%
-\nabla^{\lambda}\phi\nabla^{\rho}\phi R_{\mu\lambda\nu\rho}\\
&&-(\nabla_{\mu}\nabla^{\lambda}\phi)(\nabla_{\nu}\nabla_{\lambda}%
\phi)+(\nabla_{\mu}\nabla_{\nu}\phi)\square\phi+\frac{1}{2}G_{\mu\nu}%
(\nabla\phi)^{2}\\
&& -g_{\mu\nu}\left[  -\frac{1}{2}(\nabla^{\lambda}\nabla^{\rho}\phi
)(\nabla_{\lambda}\nabla_{\rho}\phi)+\frac{1}{2}(\square\phi)^{2}%
-\nabla_{\lambda}\phi\nabla_{\rho}\phi R^{\lambda\rho}\right]  \ ,\\
T_{\mu\nu}^{em}  &=&F_{\mu}{}^{\lambda}F_{\nu\lambda}-\frac{1}{4}g_{\mu\nu
}F^{2}\ .
\end{eqnarray*}
We will consider the family of spacetimes
\begin{equation}
ds^{2}=-F(r)dt^{2}+G(r)dr^{2}+r^{2}d\Sigma_{K}^{2}\ , \label{metric}%
\end{equation}
where $d\Sigma_{K}$ is the line element of a closed, $(n-2)$-dimensional
Euclidean space of constant curvature $K=0,\pm1$.\ The metric (\ref{metric})
corresponds to the most general static spacetime compatible with the possible
local isometries of $\Sigma_{K}$ acting on a spacelike section. For $K=1$, the
space $\Sigma_{K}$ is locally a sphere, for $K=0$ it is locally flat, while
for $K=-1$ it locally reduces to the hyperbolic space. Hereafter we will
consider a static and isotropic scalar field, i.e. $\phi=\phi\left(  r\right)
$.

The outline of the paper is as follows: in section 2 the four-dimensional
solution is given for arbitrary $K$, and the energy density is computed. In
section 3, the spherically symmetric solution is described in detail and the
constraints in the couplings parameters are described in order to obtain a
real scalar field and positive energy density. We comment as well on some of
the thermodynamical properties of the solution. In section 4, the solution in
arbitrary dimension $n$ is given. Finally in section 5 the solution in the
special case when $\alpha=0$ is analyzed. In this paper we use the ``mostly
plus signature" and Greek indices stand for indices in the coordinate basis.

\section{Four dimensional solution}

Using the ansatz (\ref{metric}) the equation of motion for the scalar field
(\ref{eqphi}) admits a first integral, which implies the equation
\begin{equation}
r\frac{F^{\prime}(r)}{F(r)}=\left[  K+\frac{\alpha}{\eta}r^{2}-\frac{C_{0}%
}{\eta}\frac{G(r)}{\psi(r)\sqrt{F(r)G(r)}}\right]  G(r)-1\ ,
\label{eqfieldfixed}%
\end{equation}
where $C_{0}$ is an integration constant, $\psi(r):=\phi^{\prime}(r)$, and
$(^{\prime})$ stands for derivation with respect to $r$. As it was done in
reference \cite{rinaldisolo}, and then in \cite{ado} we (arbitrarily) set
$C_{0}=0$, which allows to find a simple relation between the metric
functions $F(r)$ and $G(r)$
\begin{equation}
G(r)=\frac{\eta}{F(r)}\left(  \frac{rF^{\prime}(r)+F(r)}{r^{2}\alpha+\eta
K}\right)  \ . \label{G(r)}%
\end{equation}
The Maxwell equation admits a first integral as well,
providing the following relation
\begin{equation}
G(r)=\frac{r^{4}}{q^{2}F(r)}(A_{0}^{\prime}(r))^{2}\ ,
\end{equation}
where $\frac{1}{q^{2}}$ is an integration constant. These two last equations
allow us to find an expression for the first radial derivative of the electric
potential
\begin{equation}
(A_{0}^{\prime}(r))^{2}=\frac{q^{2}\eta}{r^{4}}\left(  \frac{rF^{\prime
}(r)+F(r)}{r^{2}\alpha+\eta K}\right)  \ . \label{A(r)}%
\end{equation}

In this way, equations (\ref{G(r)}) and (\ref{A(r)}) together with the $tt$
and $rr$ components of (\ref{eqmetric}), provide a consistent system which for
$K=\pm1$ and $\eta\Lambda\neq\alpha$, has the following solution
\begin{align*}
F(r)  &  =\frac{r^{2}}{l^{2}}+\frac{K}{\alpha}\sqrt{\alpha\eta K}\left(
\frac{\alpha+\Lambda\eta+\frac{\alpha^{2}}{4\eta\kappa K}q^{2}}{\alpha
-\Lambda\eta}\right)  ^{2}\frac{\arctan\left(  \frac{\sqrt{\alpha\eta K}}{\eta
K}r\right)  -\mu}{r}\\
&  +\frac{\alpha^{2}}{\kappa(\alpha-\Lambda\eta)^{2}}\frac{q^{2}}{r^{2}}%
+\frac{\alpha^{3}}{16\eta\kappa^{2}K^{2}(\alpha-\Lambda\eta)^{2}}\frac{q^{4}%
}{r^{2}}-\frac{\alpha^{2}}{48\kappa^{2}K(\alpha-\Lambda\eta)^{2}}\frac{q^{4}%
}{r^{4}}+\frac{3\alpha+\Lambda\eta}{\alpha-\Lambda\eta}K\ ,\\[10pt]
G(r)  &  =\frac{1}{16}\frac{\alpha^{2}(4\kappa\left(  \alpha-\eta
\Lambda\right)  r^{4}+8\eta\kappa Kr^{2}-\eta q^{2})^{2}}{r^{4}\kappa
^{2}(\alpha-\eta\Lambda)^{2}(\alpha r^{2}+\eta K)^{2}F(r)}\ ,\\[10pt]
\psi^{2}(r)  &  =-\frac{1}{32}\frac{\alpha^{2}(4\kappa(\alpha+\eta
\Lambda)r^{4}+\eta q^{2})(4\kappa\left(  \alpha-\eta\Lambda\right)
r^{4}+8\eta\kappa Kr^{2}-\eta q^{2})^{2}}{r^{6}\eta\kappa^{2}(\alpha
-\eta\Lambda)^{2}(\alpha r^{2}+\eta K)^{3}F(r)} \ ,\label{psisol}\\[10pt]
A_{0}(r)  &  =\frac{1}{4}\frac{q\sqrt{\alpha}}{\eta^{\frac{3}{2}}K^{\frac
{5}{2}}\kappa}\left(  \frac{4\beta\kappa K^{2}(\alpha+\eta\Lambda)+\alpha
^{2}q}{(\alpha-\eta\Lambda)}\right)  \arctan\left(  \frac{\sqrt{\alpha\eta K}%
}{\eta K}r\right) \\
&  +\alpha\left(  \frac{8\eta\kappa K^{2}+\alpha q}{4\eta\kappa K^{2}%
(\alpha-\eta\Lambda)}\right)  \frac{q}{r}-\frac{\alpha}{12\kappa K(\alpha
-\eta\Lambda)}\frac{q^{3}}{r}\ .
\end{align*}

Here we have defined the effective (A)dS radius $l$ by $l^{-2}:=\frac{\alpha
}{3\eta}$. In the case of a locally flat transverse section ($K=0$) the system
integrates in a different manner and the solution takes the form
\begin{align*}
F(r)  &  =\frac{r^{2}}{l^{2}}-\frac{\mu}{r}+\frac{\alpha}{2\kappa(\alpha
-\eta\Lambda)}\frac{q^{2}}{r^{2}}+\frac{\alpha\eta}{80\kappa^{2}(\alpha
-\eta\Lambda)^{2}}\frac{q^{4}}{r^{6}}\ ,\\
G(r)  &  =\frac{1}{16}\frac{\left(  4\kappa(\alpha-\eta\Lambda)r^{4}-\eta
q^{2}\right)  ^{2}}{\kappa^{2}(\alpha-\Lambda\eta)r^{8}F(r)}\ ,\\
\psi(r)^{2}  &  =-\frac{1}{32}\frac{\left(  4\kappa(\alpha+\eta\Lambda
)r^{4}+\eta q^{2}\right)  (4\kappa(\alpha-\eta\Lambda)r^{4}+\eta q^{2})^{2}%
}{\alpha\eta r^{12}\kappa^{2}(\alpha-\Lambda\eta)^{2}F\left(  r\right)  }\ ,\\
A_{0}(r)  &  =-\left(  \frac{20\kappa(\alpha-\eta\Lambda)r^{4}-\eta q^{2}}
{20\kappa(\alpha-\eta\Lambda)r^{5}}\right)  q\ .
\end{align*}

In the case when we set $q\rightarrow0$ we recover the result obtained in
\cite{ado} for the cases $K=\pm1$ as well as for the case $K=0$. The later case
reduces to topological Schwarzschild solution with locally flat horizon
\cite{Lemos}.

It can be seen that this solution is asymptotically locally dS or AdS for
$\alpha/\eta<0$ or $\alpha/\eta>0$, respectively, since when $r\rightarrow
\infty$ the components of the Riemann tensor go to
\[
R_{\ \ cd}^{ab}\underset{r\rightarrow\infty}{=}-\frac{\alpha}{3\eta}%
\delta_{cd}^{ab}:=-\frac{1}{l^{2}}\delta_{cd}^{ab}\ ,
\]
justifying our previous definition of the effective (A)dS radius. The
asymptotic expansion ($r\rightarrow\infty$) of the metric functions and of the
gauge field reads
\begin{eqnarray*}
g_{tt}&\underset{r\rightarrow\infty}{=}&\frac{r^{2}}{l^{2}}+\frac
{3\alpha+\eta\Lambda}{\alpha-\eta\Lambda}K+\frac{K}{2\alpha}\sqrt{\alpha\eta
K}\left(  \frac{(\alpha+\eta\Lambda)+\frac{\alpha^{2}q^{2}}{4\eta\kappa K^{2}%
}}{\alpha-\eta\Lambda}\right)  ^{2}\frac{\pi\sigma-2\mu}{r}+O\left(
r^{-2}\right)  \ ,\\
g^{rr}&\underset{r\rightarrow\infty}{=}&\frac{r^{2}}{l^{2}}+\frac
{7\alpha+\eta\Lambda}{3(\alpha-\eta\Lambda)}K+\frac{K}{2\alpha}\sqrt
{\alpha\eta K}\left(  \frac{(\alpha+\eta\Lambda)+\frac{\alpha^{2}q^{2}}%
{4\eta\kappa K^{2}}}{\alpha-\eta\Lambda}\right)  ^{2}\frac{\pi\sigma-2\mu}%
{r}+O\left(  r^{-2}\right)  \ ,\\
A_{0}(r)&\underset{r\rightarrow\infty}{=}&a_{0}-\frac{q}{r}+O(r^{-2})\ ,
\end{eqnarray*}

\bigskip

\noindent where $\sigma$ is the sign of $\eta K$ and $a_{0}$ is  a constant. From here it is
possible to see that our electric potential reproduces the Coulomb potential at
infinity. There is a curvature singularity at $r=0$ since for example the
Ricci scalar diverges as
\begin{equation}
R\underset{r\rightarrow0}{=}\frac{4K}{r^{2}}+O(1)\ .
\end{equation}

If $\rho(r)$ is the energy density, then the total energy $\mathcal{E}$\ is
given by
\begin{equation}
\mathcal{E}=V\left(  \Sigma\right)  \int dr\rho\left(  r\right)  \ ,
\end{equation}
where $V\left(  \Sigma\right)$ stands for the volume of $\Sigma$.
Therefore
\begin{equation}
\rho\left(  r\right)  :=r^{2}\sqrt{G\left(  r\right)  }F\left(  r\right)
^{-1}T_{tt}\ .
\end{equation}
Now, the $tt$ component of the energy momentum tensor reads
\begin{equation}
T_{tt}=-\frac{(\alpha+\Lambda\eta)}{\eta\kappa^{2}}F(r)\left[
1-H(r)F(r)\right]  \ ,
\end{equation}
where $H(r)$ is the given by the expression
\[
H(r)=\frac{64\eta^{2}r^{2}(\alpha-\Lambda\eta)^{2}(r^{2}\alpha+\eta K)}%
{\alpha^{2}\kappa^{2}(\alpha+\Lambda\eta)}\left(  \frac{q^{2}\kappa
(2r^{2}\alpha+\eta K)-4 K(\alpha+\Lambda\eta)r^{4}}{4(\alpha-\Lambda\eta
)r^{4}+8 r^{2}\eta K-\eta\kappa q^{2}}\right)  \ .
\]
\noindent If we take the limit $q\rightarrow0$ we recover the $T_{tt}$
component of the uncharged case.

\section{Spherically symmetric case}

Now we study the particular case with a spherically symmetric transverse
section $K=1$. The solution for the metric components and for the square of
the derivative of the scalar field reduces to
\begin{align*}
F(r)  &  =\frac{r^{2}}{l^{2}}+\frac{1}{\alpha}\sqrt{\alpha\eta}\left(
\frac{\alpha+\Lambda\eta+\frac{\alpha^{2}}{4\eta\kappa}q^{2}}{\alpha
-\Lambda\eta}\right)  ^{2}\frac{\arctan\left(  \frac{\sqrt{\alpha\eta}}{\eta
}r\right)  -\mu}{r}\\[5pt]
&  +\frac{\alpha^{2}}{\kappa(\alpha-\Lambda\eta)^{2}}\frac{q^{2}}{r^{2}}%
+\frac{\alpha^{3}}{16\eta\kappa^{2}(\alpha-\Lambda\eta)^{2}}\frac{q^{4}}%
{r^{2}}-\frac{\alpha^{2}}{48\kappa^{2}(\alpha-\Lambda\eta)^{2}}\frac{q^{4}%
}{r^{4}}+\frac{3\alpha+\Lambda\eta}{\alpha-\Lambda\eta}\ ,\\[10pt]
G(r)  &  =\frac{1}{16}\frac{\alpha^{2}(4\kappa\left(  \alpha-\eta
\Lambda\right)  r^{4}+8\eta\kappa r^{2}-\eta q^{2})^{2}}{r^{4}\kappa
^{2}(\alpha-\eta\Lambda)^{2}(\alpha r^{2}+\eta)^{2}F(r)}\ ,\\[10pt]
\psi^{2}(r)  &  =-\frac{1}{32}\frac{\alpha^{2}(4\kappa(\alpha+\eta
\Lambda)r^{4}+\eta q^{2})(4\kappa\left(  \alpha-\eta\Lambda\right)
r^{4}+8\eta\kappa r^{2}-\eta q^{2})^{2}}{r^{6}\eta\kappa^{2}(\alpha
-\eta\Lambda)^{2}(\alpha r^{2}+\eta)^{3}F(r)}\ ,\\[10pt]
A_{0}(r)  &  =\frac{1}{4}\frac{q\sqrt{\alpha}}{\eta^{\frac{3}{2}}\kappa
}\left(  \frac{4\beta\kappa^{2}(\alpha+\eta\Lambda)+\alpha^{2}q}{(\alpha
-\eta\Lambda)}\right)  \arctan\left(  \frac{\sqrt{\alpha\eta}}{\eta}r\right)
\\[5pt]
&  +\alpha\left(  \frac{8\eta\kappa+\alpha q}{4\eta\kappa(\alpha-\eta\Lambda
)}\right)  \frac{q}{r}-\frac{\alpha}{12\kappa(\alpha-\eta\Lambda)}\frac{q^{3}%
}{r}\ .
\end{align*}

In order to analize the features proper of a black hole in our solution we
need to analize the lapse function $F(r)$. As we approach the origin, the
lapse function goes to minus infinity. On the other hand, as we go to
infinity along coordinate $r$, $F(r)$ tends to plus infinity. Therefore, it is
clear that this function being continuous has at least one cero. We can prove
that this function has more than one cero. Since we know the existence of at
least one cero $r_{H}$, we can parametrize the function with $r_{H}$ as
parameter. From the equation $F(r_{H})=0$ we get $\mu\equiv\mu(r_{H})$ which can be used
to express the lapse function as $F(r,\mu(r_{H}))$. To prove the existence of
the second event horizon, we can do the same as before but with the electric
charge. We propose the existence of $r_{h}$, then $F(r_{h})=0$, and using this
we get $q^{2}\equiv q^{2}(r_{h},r_{H})$. It is possible to find two roots for
$F(r_{h})=0$ or in other words, two suitable values of $q^{2}$ for a possible
$r_{h}$. This values in some cases are both negatives, both positive or one
positive and the second negative, but at least the existence of one positive
root is enough to prove the existence of $r_{h}$. As we said, due to the shape
near the origin and at infinity of the lapse function, the existence of two
zeros of the function implies the existence of a third zero for some range of
paramaters. Therefore $F(r)$ can have just one zero, two zeros\footnote{This
case is an special case in the sense that contains a zero which is a local
minimum. When that local minimum is the outer horizon this corresponds to an
extremal black hole.} or three zeros. Each of these cases exist for a specific
set of values of the coupling and cosmological constants. From hereafter and
for simplicity, we will focus in the case when the lapse function has just one zero.

Reality condition of the lapse function requires $\alpha \eta>0$. Therefore $l^{-2}:=\frac{\alpha}{3\eta}$ is positive defined and
the spacetime is asymptotically AdS. As it was noted in the uncharged case
\cite{ado} without loss of generality it is possible to choose both parameters
positive, since the solution with both $\alpha$ and $\eta$ negative is
equivalent to the former by changing $\mu\rightarrow-\mu$.

In order to obtain a real scalar field in the domain of outer communications
and satisfy the positivity of the energy, we need to impose some constraints
in our parameters. In fact, the value of the cosmological constant is
restricted to be
\begin{equation}
\label{restr}\Lambda<-\frac{q^{2}}{4r_{H}^{4}\kappa}-\frac{\alpha}{\eta}\ .
\end{equation}
It is important to note that we cannot switch off the scalar field. This
implies that our solution is not continuously connected with the maximally
symmetric background. Despite of this, setting $\mu=0$ and $q=0$ we observe
that the spacetime is regular, actually is the only regular spacetime that can
be found within this family. Such a case describes an asymptotically AdS
gravitational soliton. Close to $r=0$ and after a proper reescaling on the
time coordinate the spacetime metric takes the following form
\begin{equation}
ds_{soliton}^{2}=-\left(  1-\frac{\Lambda}{3}r^{2}+O(r^{4})\right)
dt^{2}+\left(  1-\frac{3\alpha+2\Lambda\eta}{3\eta}r^{2}+O(r^{4})\right)
dr^{2}+r^{2}d\Omega^{2}\ .
\end{equation}
The thermal version of this spacetime can be used as the background metric for
obtain a regularized euclidean action which could be used to obtain the
thermodynamical properties of the black holes in the Hawking-Page approach.

\section{N-dimensional case}

In this section we analize the $n-$dimensional solution to the action
principle defined by (\ref{action}). For doing this, we take the variation of
our Lagrangian with respect to all the functions involved $F(r)$, $G(r)$,
$\phi(r)$ and $A_{0}(r)$. This procedure gives us the equations of motion of
the system.

Therefore, following the same strategy than in four dimensions, the equation of motion for the scalar field
admits a first integral. Setting to zero the integration constant of this
equation we obtain a relation between the metric coefficients, but now in
arbitrary dimension
\begin{equation}
G_{n}(r)=\frac{\eta(n-2)}{F_{n}(r)}\left(  \frac{F_{n}^{\prime}(r)r+F_{n}%
(r)(n-3)}{2r^{2}\alpha+\eta K(n-2)(n-3)}\right)  \ .
\end{equation}
The equation coming from the variation with respect to the
electric field gives us the following relation
\[
\left(  A_{0_{n}}^{\prime}(r)\right)  ^{2}=q^{2}F_{n}(r)G_{n}(r)r^{(4-2n)}\ .
\]
In the same spirit, and using the last result, it is possible to obtain a
relation for $\psi(r)^{2}$. Then
\[
\psi_{n}(r)^{2}=-\frac{1}{2}(n-2)\left(  \frac{\Xi_{n}^{1}+\Xi_{n}^{2}}%
{\Xi_{n}^{3}}\right)  \ ,
\]
where we have defined
\begin{align*}
\Xi_{n}^{1}  &  =(n-3)^{2}(4\kappa\Lambda\eta r^{2}+4\kappa r^{2}\alpha
+q^{2}r^{(-2n+6)}\eta)F_{n}(r)^{2}\\
&  +2(n-3)(q^{2}r^{(-2n+7)}\eta+4\kappa\Lambda\eta r^{3}+4\alpha r^{3}%
\kappa)F_{n}^{\prime}(r)F_{n}(r)\ ,\\
\Xi_{n}^{2}  &  =(4\kappa\Lambda\eta r^{4}+q^{2}r^{(-2n+8)}\eta+4\alpha
r^{4}\kappa)F_{n}^{\prime}(r)^{2}\ ,\\
\Xi_{n}^{3}  &  =F_{n}(r)(2r^{2}\alpha\eta Kn^{2}-5\eta Kn+6\eta
K)^{2}((n-3)F_{n}(r)+F_{n}^{\prime}(r)r)\ .
\end{align*}

Using these expresions and the equation resulting from the variation with
respect to the fuction $F_{n}(r)$, we can obtain a relation which allows to
obtain the explict form of $F_{n}(r)$ for an arbitrary value of the dimension
$n$, and in this way, the complete solution to our system. We checked the
result from $n=4$ to $n=10$.

\section{Asymptotically locally flat black holes with charge supported by the
Einstein-kinetic coupling}

In this section we will study the particular case where the scalar field is
coupled to the background only with the Einstein tensor. It is possible to do
this by setting $\alpha=0$. Under the presence of an electric field, we obtain
asymptotically locally flat black hole solutions in the case where the
cosmological constant is present. Therefore, the action principle is given by
\begin{equation}
I[g_{\mu\nu},\phi]=\int\sqrt{-g}d^{4}x\left[  \kappa\left(  R-2\Lambda
\right)  +\frac{\eta}{2}G_{\mu\nu}\nabla^{\mu}\phi\nabla^{\nu}\phi-\frac{1}%
{4}F_{\mu\nu}F^{\mu\nu}\right]  \ .
\end{equation}
Following the same procedure (with $\alpha\neq0$ and
$K=1$)\footnote{In the case where $K=0$, the system integrate in a different
manner. In fact, $\Lambda$ and $q$ have to vanish in order to fulfil the field
equations. Then, we obtain the same degenerated system found in \cite{ado}.}
we obtain
\begin{equation}
ds^{2}=-F(r)dt^{2}+\frac{15[4\kappa r^{2}(2-\Lambda r^{2})-q^{2}]^{2}}{r^{4}%
}\frac{dr^{2}}{F(r)}+r^{2}d\Omega^{2}\ ,
\end{equation}
where
\begin{align*}
F(r)  &  =48\kappa^{2}\Lambda^{2}r^{4}-320\kappa^{2}\Lambda r^{2}%
+120\kappa(8\kappa+\Lambda q^{2})-\frac{\mu}{r}+240\kappa\frac{q^{2}}{r^{2}%
}-5\frac{q^{4}}{r^{4}}\ ,\\
\psi(r)^{2}  &  =-\frac{15}{2}\frac{(4\kappa\Lambda r^{4}+q^{2})(4\kappa
r^{2}(2-\Lambda r^{2})-q^{2})^{2}}{r^{6}\eta}\frac{1}{F(r)}\ ,\\
A_{0}(r)  &  =\sqrt{15}\left(  \frac{q^{3}}{3r^{3}}-8\kappa\frac{q}{r}%
-4\kappa\Lambda rq\right)  \ .
\end{align*}

This solution shows the following features:

\begin{itemize}
\item The solution is asymptotically locally flat, namely we have
\[
\lim_{r\rightarrow\infty}R^{\mu\nu}{}_{\lambda\rho}\rightarrow0\ .
\]

\item For a non degenerated horizon $r=r_{H}$ we have $F(r_{H})=0$, then the
scalar field vanishes at the horizon and is not analytic there.

\item In order to obtain a real scalar field outside of the horizon we can
impose two different conditions:

\begin{enumerate}
\item $\Lambda>0$ and $\eta<0$ or

\item $\Lambda<-\frac{q^{2}}{4\kappa r_{H}^{4}}$ and $\eta>0$.
\end{enumerate}

\item For any value of the integration constant $\mu$ we have the curvature
singularities
\begin{align*}
r_{0}  &  =0\ ,\\
r_{1,2}  &  =\frac{\sqrt{2\kappa\Lambda(2\kappa\pm\sqrt{4\kappa^{2}%
-\kappa\Lambda q^{2}})}}{2\kappa\Lambda}\ .
\end{align*}

Then for $\Lambda<0$ the only singularity is located at the
origin of coordinates. If the cosmological constant is
positive, in order to rule out the existence of singularities different than $r=0$, we need to impose the following constraint in the value of $\Lambda$%
\begin{equation}
\Lambda>\frac{4\kappa}{q^{2}}\ .
\end{equation}

\item We point out that in the limit $r\rightarrow\infty$ our electric
potential represents a constant electric field at that point supported by the
cosmological constant, and in this way we obtain an asymptotically electric Universe.

\item Finally the limit $q\rightarrow0$ we recover the results
obtained in \cite{ado}.
\end{itemize}

Let us put $\Lambda=0$, then the solution takes the form
\[
ds^{2}=-F(r)dt^{2}+\frac{3(8\kappa r^{2}-q^{2})^{2}}{r^{4}}\frac{dr^{2}}%
{F(r)}+r^{2}d\Omega^{2}\ ,
\]
\noindent where
\begin{align*}
F(r)  &  =192\kappa^{2}-\frac{\mu}{r}+48\kappa\frac{q^{2}}{r^{2}}-\frac{q^{4}%
}{r^{4}}\ ,\\
\psi(r)^{2}  &  =-\frac{15}{2}\frac{(8\kappa r^{2}-q^{2})^{2}}{r^{6}\eta}%
\frac{q^{2}}{F(r)}\ ,\\
A_{0}(r)  &  =\sqrt{15}\left(  \frac{q^{3}}{3r^{3}}-8\kappa\frac{q}{r}\right)
\ .
\end{align*}

In this case we have:

\begin{itemize}
\item The solution is asymptotically flat
\[
ds^{2}=-\left(  1-\frac{\mu}{r}+O(r^{-2})\right)  dt^{2}+\left(  1+\frac{\mu
}{r}+O(r^{-2})\right)  dr^{2}+r^{2}d\Omega^{2}\ ,
\]
which is razonable because when we have $\Lambda=0$, the electric field at
infinity vanishes.

\item For a non degenerated horizon $r=r_{H}$ we have $F(r_{H})=0$, then the
scalar field vanish at the horizon, as in the previous cases, is not analytic there.

\item In order to obtain a real scalar field outside of the horizon we impose
\[
\eta<0\ .
\]

\item For any value of the integration constant $\mu$ we have the curvature
singularities
\begin{align*}
r_{0}  &  =0\ ,\\
r_{1}  &  =\sqrt{\frac{1}{8\kappa}}|q|\ .
\end{align*}

\item The electric field goes to zero at infinity.

\item Taking the limit when $q\rightarrow0$ we obtain a trivial scalar field
and then we recover the Schwarzschild solution.
\end{itemize}

\section{Discussion}

In this work a particular sector of the Horndeski theory was considered where
the gravity part is given by the Einstein-Hilbert term, and where the matter
source is represented by a scalar field which has a non minimal kinetic
coupling constructed with the Einstein tensor. The main novelty of this work is the inclusion of the Maxwell field. We found exact solutions to this system for a spherically symmetric and topological horizons in all dimensions. The solution gives a new class of
asymptotically locally AdS and asymptotically locally flat black hole solutions.

These solutions are obtained using two important observations. The first one,
is the fact that the equation of motion for the scalar field admits a first
integral, which after setting the integration constant to zero (arbitrarily)
gives a simple relation between the two metric functions. The second one, is
that the Maxwell equations are easily integrated for our ansatz and symmetry
conditions, given a simple relation between the electric potential term and
the metric functions. Mixing these two results we obtain a complete
description of the system, obtaining in that way the exact solution for the
topological case in $n\geq4$ dimensions.

We observe and point out that in the case of the asymptotically locally AdS
solution, the cosmological constant at infinity is not given by the
cosmological $\Lambda$ term in the action but rather in terms of the coupling
constants $\alpha$ and $\eta$ that appear in the kinetic coefficients of the
field. The electric field is well behaved and goes to the Coulombian one at infinity.

The solutions are not continuously connected with the maximally symmetric AdS
or flat backgrounds since the scalar field cannot be turned off. Nevertheless,
since our family of metrics contains a further integration constant, it is
possible to show that within such a family there is a unique regular
spacetime. Such spacetime is a gravitational soliton and it is useful in the
four dimensional spherically symmetric case to define a regularized Euclidean
action and to explore the thermodynamics of the black hole solution. A similar
situation occurs with the AdS soliton, which can be considered as the
background for the planar AdS black holes, as well as in gravity in 2+1
with scalar fields, where the gravitational solitons are the right backgrounds to give a microscopic description of the black hole entropies
\cite{Correa2}\cite{Correa3}\cite{Zreview}.

In the particular case when the scalar field is only coupled to the metric
through the Einstein tensor, namely, $\alpha=0$ we obtain an asymptotically
locally flat black hole solution. When $\Lambda\neq0$ this solution presents
some interesting properties. The solution exist in both cases, where the
cosmological constant is positive and when is negative, given a real scalar
field configuration depending on constraints imposed on the electric charge
and on the coupling constant $\eta$. In any of these cases we obtain a
constant electric field at infinity, representing in this way our solution a
electric Universe. This constant electric field at infinity is just supported
by the cosmological constant.

In the case where $\Lambda=0$ we obtain a real scalar configuration just in
case where the coupling constant is negative. The solution is asymptotically
flat and the electric field vanishes at
infinity when $\Lambda=0$. If we switch off the electric field setting $q=0$,
we get a trivial scalar field and then we recover the Schwarzschild solution.

It is important to note that Horndeski theory offers the posibility of
exploring its solutions in many different ways. In another context, using the
same action principle, but without the Maxwell term an asymptotically Lifshitz
solution was recently found in \cite{mok}. Moreover, even if it is not possible
to obtain an analytic solution to the most general case of the Horndeski
theory for the general static black hole solution, it would be interesting to study the
cases where the non minimal coupling is given by more general tensors than the
Einstein one, namely the Lovelock tensors.

\bigskip

\section{Acknowledgments}

A. C. and C. E. would like to thank Andr\'{e}s Anabal\'{o}n for useful
discussions and comments. We are grateful to Julio Oliva for the useful insight during the development of this work. The work of A. C. is supported by 
CONICYT and by the project FSM1204 of Internationalization of Ph. D. programs in physical science, biotechnology and electronics from the Universidad T\'ecnica Federico Santa Mar\'ia. The work of C. E. is supported by CONICYT and Centro de Estudios Cient\'{\i}ficos (CECs)
funded by the Chilean Government through the Centers of Excellence Base
Financing Program of Conicyt.


\begin{thebibliography}{99}                                                                                               %


\bibitem {BD}C.~Brans and R.~H.~Dicke,
Phys.\ Rev.\ \textbf{124}, 925 (1961).
\bibitem{dancosm}
Daniele Bertacca, Sabino Matarrese and Massimo Pietroni [arxiv.org:0703259 [astro-ph]].

\bibitem{chungcosm}
Daniel J. H. Chung, Lisa L. Everett, Konstantin T. Matchev, [arxiv.org: 0704.3285 [hep-ph]].

\bibitem{macorracosm}
A. de la Macorra, [arxiv.org:0703702 [astro-ph]].

\bibitem {Horn1}G.~W.~Horndeski,
Int.\ J.\ Theor.\ Phys.\ \textbf{10}, 363 (1974).


\bibitem {rinaldi3}C. Deffayet, G. Esposito-Farese and A. Vikman, Phys. Rev. D
\textbf{79} (2009) 084003; C. Deffayet, S. Deser and G. Esposito-Farese, Phys.
Rev. D \textbf{80} (2009) 064015.

\bibitem {rinaldi4}C. de Rham, G. Gabadadze and A. J. Tolley, Phys. Rev. Lett.
106 (2011) 231101, C. de Rham and L. Heisenberg, Phys. Rev. D \textbf{84}
(2011) 043503.

\bibitem {rinaldi6}L. Amendola, Phys. Lett. B \textbf{301} (1993) 175.

\bibitem {rinaldi7}C. Deffayet, O. Pujolas, I. Sawicki and A. Vikman, JCAP.
\textbf{1010} (2010) 026.

\bibitem {rinaldia}J. -P. Bruneton, M. Rinaldi, A. Kanfon, A. Hees, S.
Schlogel and A. Fuzfa, [arXiv:1203.4446 [gr-qc]].

\bibitem {rinaldib}S. V. Sushkov, Phys. Rev. D \textbf{80} (2009) 103505.

\bibitem {rinaldic}C. Germani and A. Kehagias, Phys. Rev. Lett. \textbf{106}
(2011) 161302.

\bibitem {Lovelock}D.~Lovelock,
J.\ Math.\ Phys.\ \textbf{12}, 498 (1971).


\bibitem {rinaldisolo}M.~Rinaldi,
Phys.\ Rev.\ D \textbf{86}, 084048 (2012), [arXiv:1208.0103 [gr-qc]].


\bibitem {malda}J.M.Maldacena, [arXiv:9711200 [hep-th]].

\bibitem {ado}Andres Anabal\'on, Adolfo Cisterna and Julio Oliva,
[arXiv:1312.3597 [gr-qc]].

\bibitem {charm}E. Babichev and C. Charmousis, [arXiv:1312.3204 [gr-qc]].

\bibitem {portu}Masato Minamitsuji, [arxiv:1312.3759 [gr-qc]].

\bibitem {NumericalSol}T.~Kolyvaris, G.~Koutsoumbas, E.~Papantonopoulos and
G.~Siopsis,
Class.\ Quant.\ Grav. (2012), [arXiv:1111.0263 [gr-qc]].


\bibitem {Lemos}J.~P.~S.~Lemos,
Phys.\ Lett.\ B \textbf{353}, 46 (1995), [arXiv:9404041 [gr-qc]].


\bibitem {HP}S.~W.~Hawking and D.~N.~Page,
Commun.\ Math.\ Phys.\ \textbf{87}, 577 (1983).


\bibitem {Correa2}F.~Correa, C.~Martinez and R.~Troncoso,
JHEP \textbf{1202}, 136 (2012), [arXiv:1112.6198 [hep-th]].


\bibitem {Correa3}F.~Correa, A.~Faundez and C.~Martinez,
Phys.\ Rev.\ D \textbf{87}, 027502 (2013), [arXiv:1211.4878 [hep-th]].


\bibitem {Zreview}J.~Zanelli,
Class.\ Quant.\ Grav.\ \textbf{29}, 133001 (2012), [arXiv:1208.3353
[hep-th]].


\bibitem {mok}Moises Bravo-Gaete, Mokhtar Hassaine, [arXiv:1312.7736[hep-th]].
\end{thebibliography}
\end{document}